\documentclass[final,twocolumn,prl,aps,floats,amsfonts,amssymb,showpacs]{revtex4}
\usepackage{amsmath}
\usepackage{graphicx}

\begin{document}

\title{Multistability and memory effect in a highly turbulent flow: Experimental evidence for a global bifurcation}

\author{Florent Ravelet}
\author{Louis Mari{\'e}}
\author{Arnaud Chiffaudel}
\email{arnaud.chiffaudel@cea.fr}
\author{Fran\c{c}ois Daviaud}
\affiliation{Service de Physique de l'\'Etat Condens\'e, DSM, CEA Saclay, CNRS URA 2464, 
91191 Gif-sur-Yvette, France}

\date[Submitted to Phys. Rev. Lett. ]{19 May 2004}



\begin{abstract}

We report an experimental evidence of a global bifurcation on a highly
turbulent von K{\'a}rm{\'a}n flow. The mean flow presents multiple
solutions: the canonical symmetric solution becomes marginally unstable
towards a flow which breaks the basic symmetry of the driving apparatus
even at very large Reynolds number. The global bifurcation between these states is
highly subcritical and the system thus keeps a memory of its history.
The transition recalls low-dimension dynamical systems transitions
and exhibits a very peculiar statistics. We discuss the role of
turbulence in two ways: the multiplicity of hydrodynamical solutions and
the effect of fluctuations on the nature of transitions.

\end{abstract}

\maketitle


Non-linear systems generally present multiple solutions and various
transitions between them. Moreover, stability and transitions are
influenced by the presence of noise and/or fluctuations. In the field of
turbulence, the question of multistability of turbulent flows, for
example in tornadoes \cite{burfos77,shthus93}, delta wing flow
\cite{gomzak85}, wakes \cite{schewe83}, and vortex breakdown
\cite{shthus99}, remains open and unsolved. While multiple analytical or
numerical solutions are often encountered for the Navier-Stokes equation
at even moderate Reynolds number (e.g., for swirling flows
\cite{batch51,stew53,zandij87,shthus93,shthus99}), these solutions are
generally neither experimentally relevant, nor stable at very high
Reynolds number. Furthermore, turbulent flows at very high Reynolds
number are generally expected to statistically respect the basic
symmetries of their driving apparatus. Indeed, even if bifurcations and
symmetry breaks occur on the way to turbulence, the fully developed
turbulent state is known to restore the broken symmetries, in the limit of infinite Reynolds number and far from boundaries \cite{frisch}.
In this Letter, we experimentally study the multistability of a
turbulent von K{\'a}rm{\'a}n flow between two counter-rotating disks in
a finite vessel at very high Reynolds number. This system undergoes a
subcritical global bifurcation between turbulent states characterized by
mean flows of different topology and symmetry. These turbulent states
coexist at high Reynolds number and can be ``prepared'' specifically,
i.e., they keep a memory of the system history. Since these states are
highly fluctuating turbulent states, we also address the question of the
role of the fluctuations for such a transition. Actually, the effect of
an external noise on an existing transition is well documented
\cite{faraday}, but the global bifurcation reported here does only take
place over an already fluctuating turbulent regime. Do fluctuations
trigger the bifurcation as multiplicative noise do for nonlinear
oscillators \cite{malmar03} and turbulent $\alpha$-effect do for dynamo
action \cite{moffatt} ? 

\paragraph*{Experimental setup.} 

\begin{figure}[!b] 
\begin{center}
\includegraphics[clip,height=2cm]{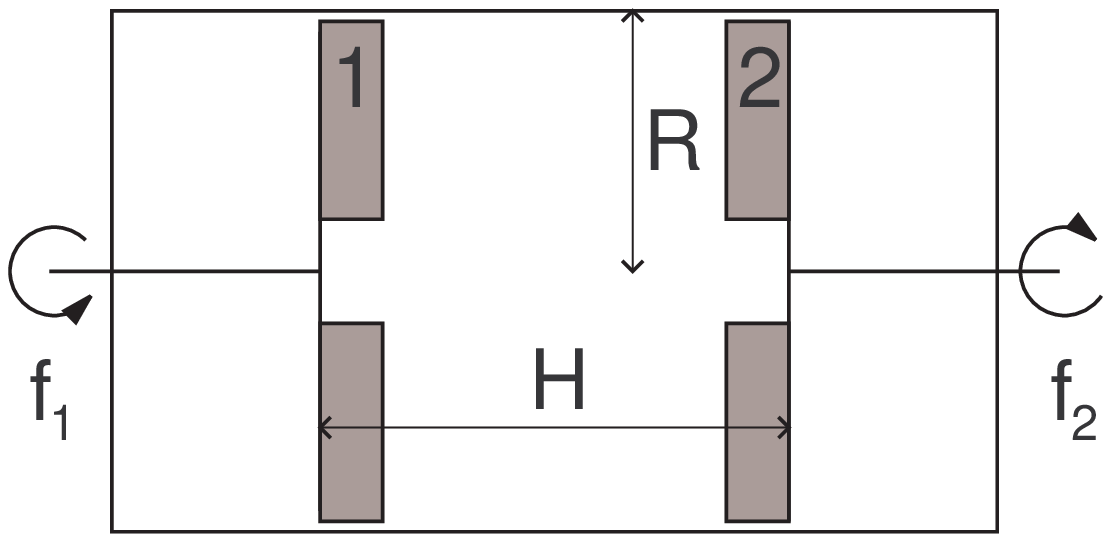}
\hspace{0.7cm}
\includegraphics[clip,height=2cm]{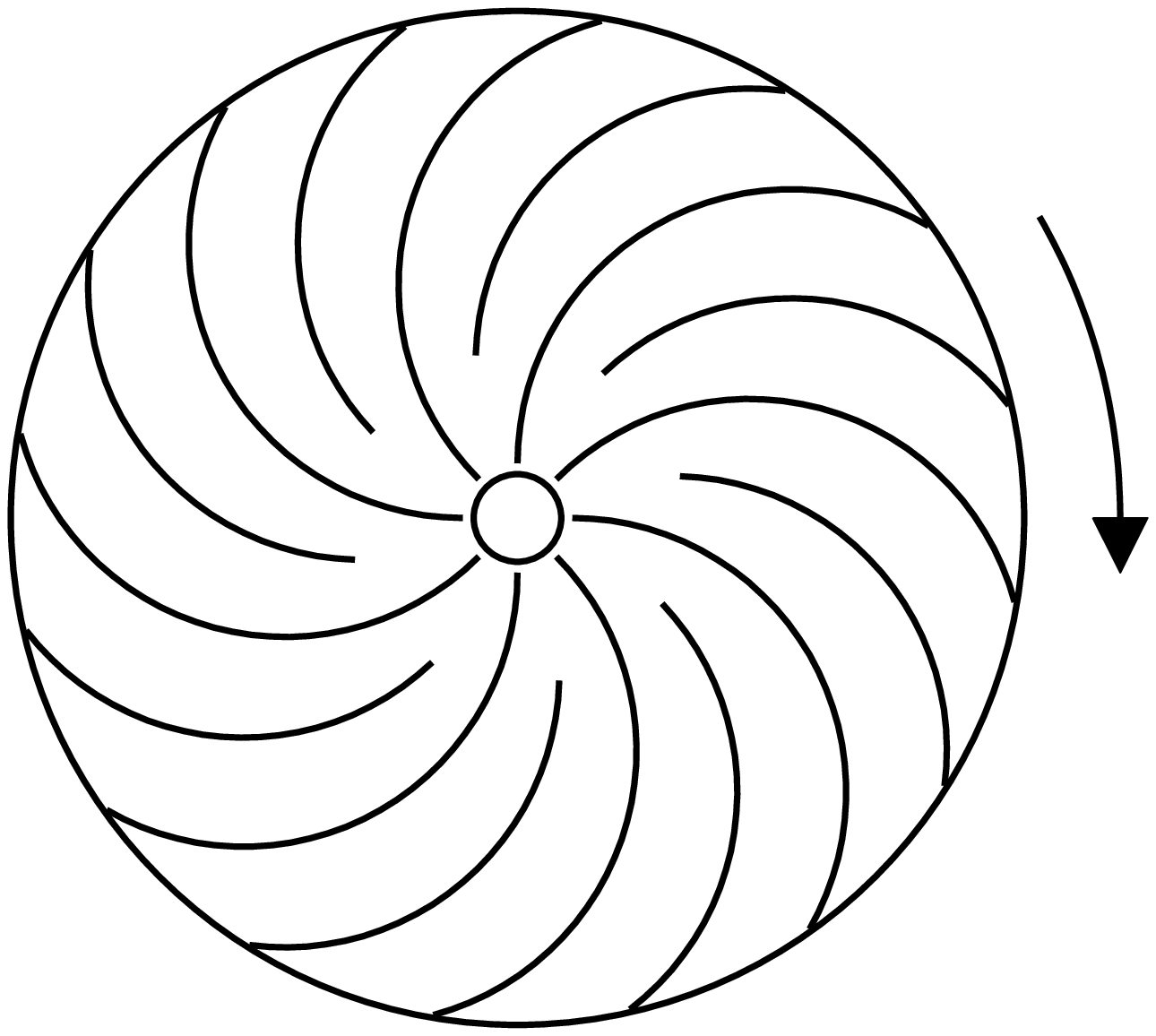} 
\end{center} 
\caption{Sketch of the experimental setup and of the impellers blades profile. 
The arrow indicates the positive rotation sense.}
\label{fig:schema} 
\end{figure}

We call von K{\'a}rm{\'a}n type flow the flow generated between two coaxial
counter-rotating impellers in a cylindrical vessel. The cylinder radius
and height are respectively $R=100$ mm and $H_c=500$ mm. We use bladed disks to ensure inertial stirring. 
Most of the inertially driven von K{\'a}rm{\'a}n setups studied in
the past dealt with straight blades \cite{caddou95,labpin96}. In this Letter, the impellers
consist of $185$ mm diameter disks each fitted with $16$ curved
blades ---curvature radius $50$ mm, height $20$ mm (Fig. \ref{fig:schema}). The
distance between the inner faces of the disks is $H=180$ mm which
defines a working space for the flow of aspect ratio ${H}/{R}=1.8$.
With curved blades, the directions of rotation are no longer equivalent.
We rotate the impellers clockwise (with the concave face
of the blades). Four baffles ($10 \times 10 \times 125$ mm) can be
added along the cylinder wall.

The impellers are driven by two independent brush-less $1.8$ kW motors, with a speed servo loop control. The motor rotation
frequencies $f_1$, $f_2$ can be varied independently in the range $0 -
15$ Hz. An experiment is thus characterized by two numbers:
$f=\sqrt{{(f_1^2+f_2^2)}/2}$ measuring the intensity of the
forcing and $\theta={(f_2-f_1)}/{(f_1+f_2)}$ measuring the
speed dissymmetry ($-1\leq\theta\leq1$). For exact counter rotation
$f_1=f_2=f$ and $\theta=0$. The speed servo loop control ensures a
precision of $0.5 \%$ on $f$, and an absolute precision of $\pm 0.002$
on $\theta$ for small values.

The working fluid is water. Copper cooling coils behind the impellers
and a thermoregulated bath ensure a thermal regulation with a precision of
$1^o$C. Velocity fields are measured
by Laser Doppler Velocimetry (LDV). Torques are measured as an image of the
current consumption in the motors given by the servo
drives and have been calibrated by calorimetry. The analog
signal is low-pass filtered at $10$ Hz. For a typical frequency $f=4$ Hz at $35^o$C, the
integral Reynolds number is $Re={2 \pi f R^2}{\nu}^{-1} \simeq 3.
\;10^5$ and the velocity fluctuation level is of order $30\%$: the flow is highly turbulent.

The von K{\'a}rm{\'a}n flow phenomenology is the
following. Each impeller acts as a centrifugal pump: the fluid rotates
with the impeller and is expelled radially. It is pumped in the center of
the impeller. In the exact counter-rotating regime, the flow is divided
into two toric cells separated by an azimuthal shear layer. The problem (equation
and boundary conditions) is invariant under rotations of $\pi$ ($\cal R
_{\pi}$) around any radial axis passing through the center of the
cylinder. The velocity field is expected $\cal R_{\pi}$-invariant.

\paragraph*{A ``statistical'' symmetry breaking.}

\begin{figure}[!b] 
\begin{center}
\includegraphics[clip,height=4.4cm]{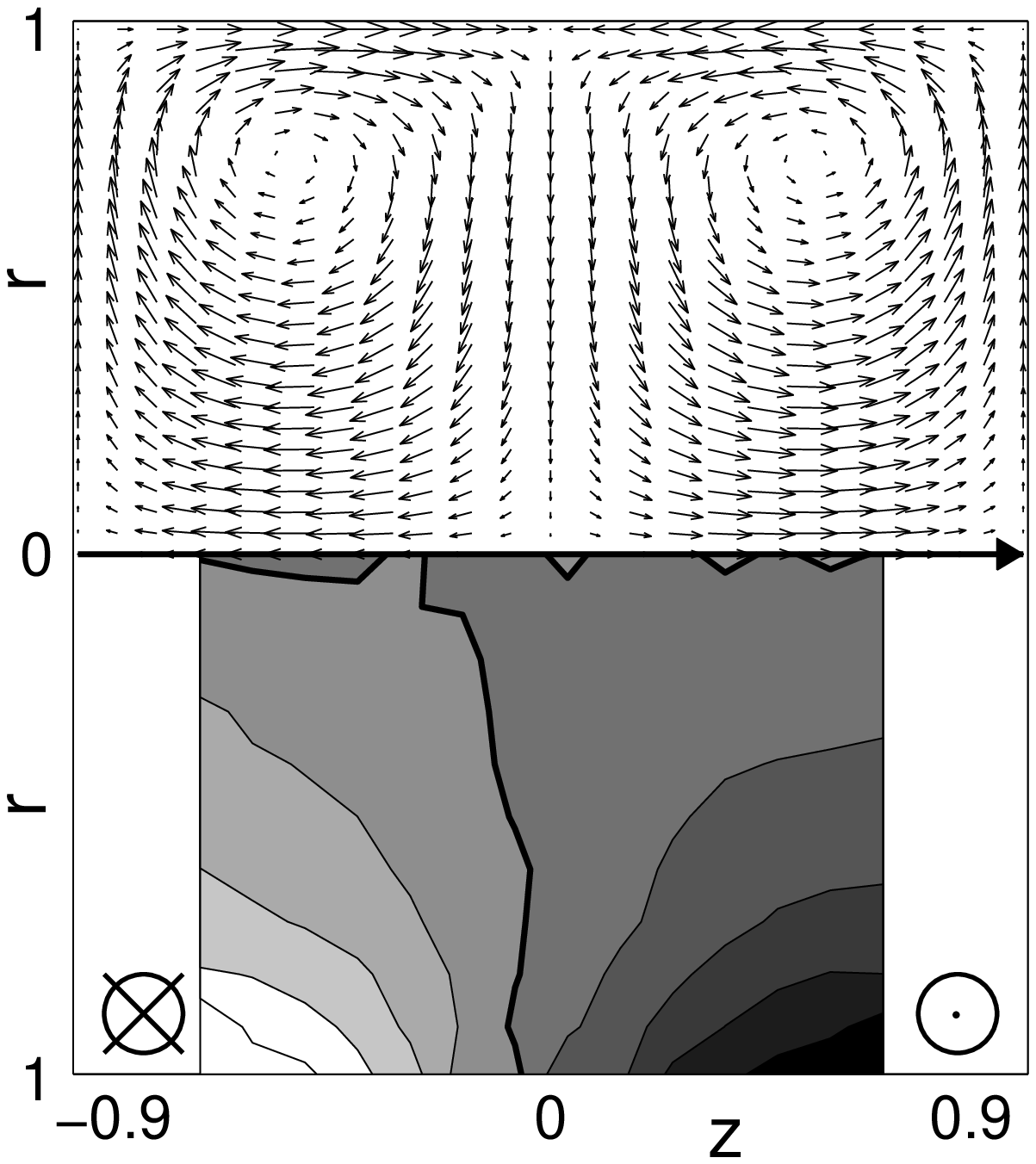}%
\includegraphics[clip,height=4.4cm]{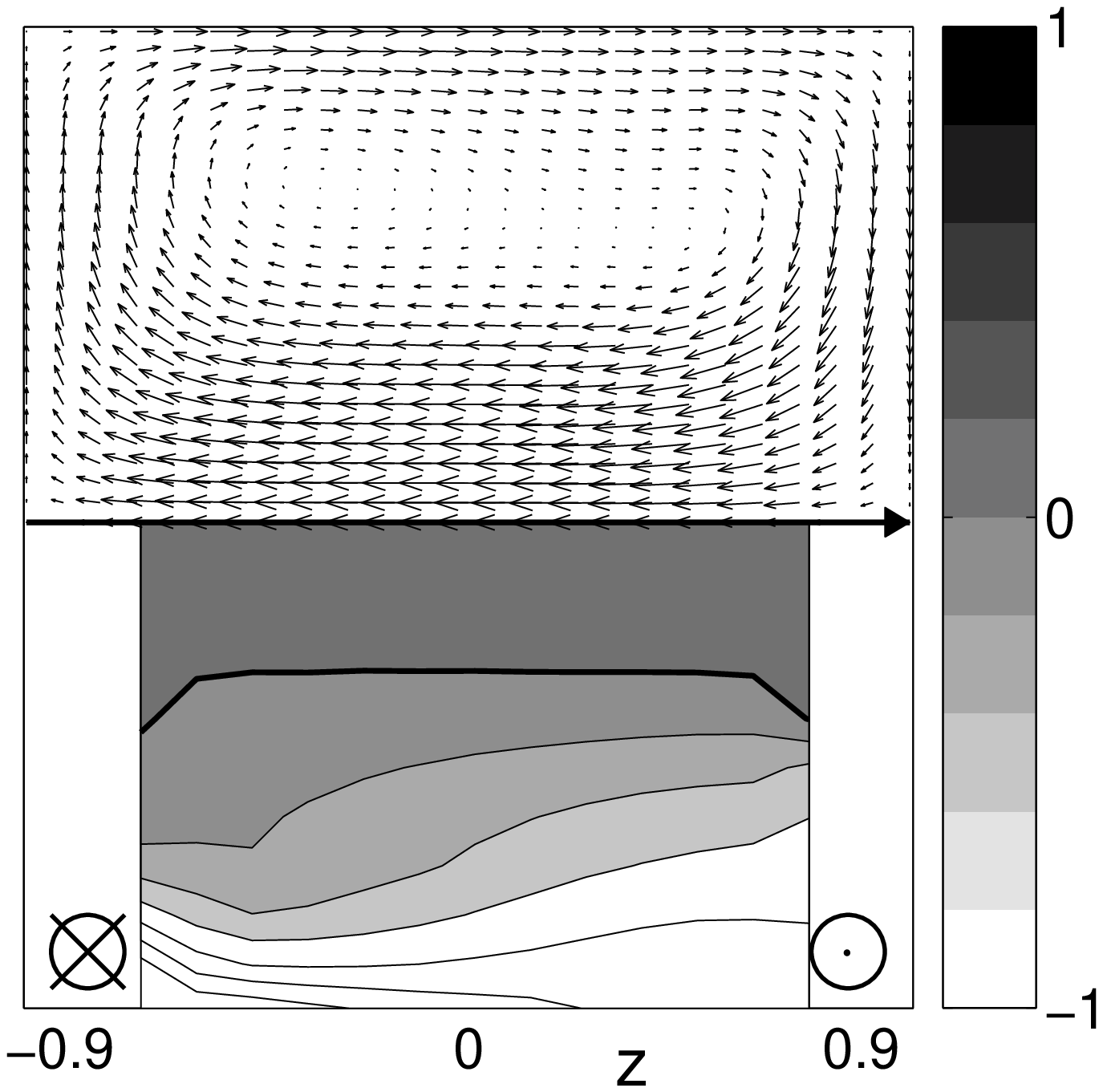}
\end{center} 
\caption{Dimensionless mean velocity field measured at $\theta=0$ by LDV over $120$ integral
turn-over time by grid point to ensure good convergence; $f=2 \;$Hz
($Re=1.5 \; 10^5$). Left: symmetric state ($s$). Right:
bifurcated state ($b_1$). Space coordinates in units of $R$. Gray code stands for azimuthal velocity.  Isolines are distant of $0.2$ and the gray code saturates in the right map. Bold lines indicate level zero.}
\label{fig:ldv1} 
\end{figure}

In our high Reynolds number regime, the flow is highly turbulent. For
instance the $rms$ value of the velocity is of the same order of
magnitude as the mean value. In Fig. \ref{fig:ldv1}~(left), we present a
map of the mean part of the exact counter-rotation flow measured by
LDV. Two cells are observed, the flow is $\cal R_{\pi}$-invariant: the
symmetries are statistically restored \cite{frisch}. The
mean angular momentum of the fluid is equal to zero: the two
impellers produce the same mean torque to maintain the flow. This situation
is well-known and documented. We label this symmetric state ($s$).

\begin{figure}[!b] 
\begin{center}
\includegraphics[clip,width=7.5cm]{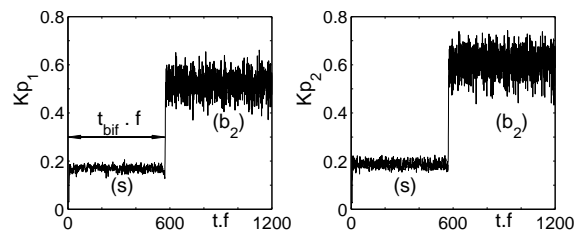}%
\end{center} 

\caption{Time serie of dimensionless torque showing the bifurcation ($s$) $\rightarrow$ ($b_2$), for $\theta=0.0204$, $f= 4.08 \;$Hz.
Left: torque on impeller~$1$. Right: torque on impeller~$2$. The bifurcation time is the time when the
torque on impeller~$1$ reaches $140\%$ of the mean value for the
symmetric state ($s$).}

\label{fig:bif} 
\end{figure}

However, with our curved blades, we observe for small $\theta$ a
global bifurcation of the flow after a certain time $t_{bif}$: both mean
velocity field and torques display dramatic changes (Fig.
\ref{fig:bif}). The two torques are suddenly $4$ times larger, and are
no longer equal. The mean flow exhibits only one cell (Fig.
\ref{fig:ldv1}, right). In the bulk, the fluid is pumped toward
impeller $1$ without rotation. Then the fluid is expelled radially and
starts spiraling along the cylinder until it meets impeller $2$ which
rotates in the opposite direction. It is abruptly stopped and reinjected
near the axis. We label this state ($b_1$). A third state ($b_2$) is
deduced from ($b_1$) by exchanging the roles of impellers $1$ and $2$.
In bifurcated states ($b_1$) or ($b_2$), the fluid is globally in
rotation: the mean angular momentum is not zero.  

Finally, three states are observed: the canonical $\cal R_{\pi}$-invariant---in a
statistical sense---state ($s$) and two bifurcated states which break the $\cal R_{\pi}$ symmetry at $\theta=0$, but are images one of the other by $\cal R_{\pi}$. We detail in the next
section the transitions between these different states.

\paragraph*{Hysteresis loops.} 

The difference between the two torques characterizes the different
states. We have checked that, as expected for so high a Reynolds number
\cite{frisch}, the torque $T$ given by one impeller for a given $(f ,
\theta)$ does not depend on $Re$ and scales as: $T(f,\theta) \; = \; K_p(\theta)
\; \rho R^5 \; (2 \pi f)^2$ \cite{labpin96}, with $\rho$ the fluid density
and $K_p$ a dimensionless power coefficient. 

\begin{figure}[!b] 

\begin{center}
\includegraphics[clip,width=8.5cm]{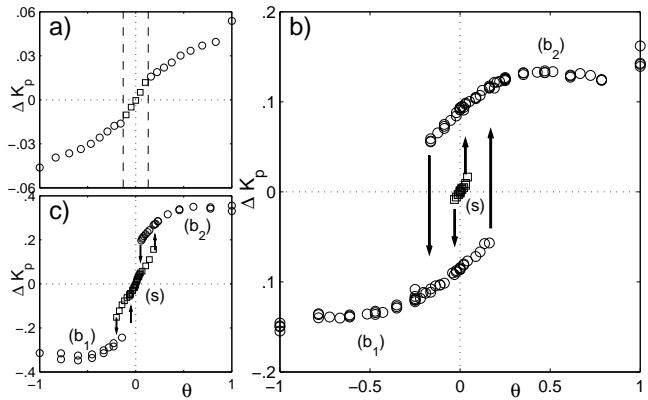}%
\end{center} 

\caption{Dimensionless torque difference $\Delta K_p$ {\it vs.}
$\theta$ for  $Re$ in the range $2-8\times10^5$. Straight blades (a) exhibit continuous transition from
$1$-cell flow to $2$-cells flow for $\theta=\pm0.13$ (vertical lines).
Curved blades without (b) or with (c) baffles along cylinder wall show
subcritical transitions between symmetric/$2$-cells ($s$)-($\Box$) and bifurcated/$1$-cell ($b_1$)-($b_2$)
states ($\bigcirc$).
}

\label{fig:loop} 
\end{figure}

In Fig. \ref{fig:loop}, we plot the dimensionless difference $\Delta
K_p$ between the two torques {\it vs.} $\theta$ for several configurations.
For straight blades, we observe a continuous curve from $\theta=-1$ to
$\theta=1$ (Fig. \ref{fig:loop}a) with two transitions between one- and
two-cells flows at $\theta=\pm0.13$.
For impellers with curved blades and no baffles on the cylinder wall, we
observe the three states in Fig. \ref{fig:loop}b. For $\theta=0$, we
recognize state ($s$) ($\Delta K_p =0$), and both bifurcated states
($b_1$) and ($b_2$). State ($s$) branch is almost reduced to one point 
and can only be reached by starting the two motors simultaneously. Its
stability is discussed in the next section.
The bifurcated state ($b_1$) lies on a branch coming continuously from
$\theta=-1$ ($f_2=0$). Starting from $\theta=-1$ and increasing
$\theta$, we stay on the ($b_1$) branch even for $\theta>0$: impeller $1$
keeps rotating and pumping the fluid although its rotation rate is
weaker than impeller $2$ rotation rate. For $\theta \simeq 0.16$ there is a
transition from ($b_1$) to ($b_2$): the fluid abruptly changes its sense of
rotation. There is a large hysteretic cycle. Note that it is impossible to reach the symmetric state ($s$) by
this way. The
global quantities of this highly turbulent flow keep memory of the way the
system has been started from rest.
An intermediate situation is reached with the same curved blades and
baffles on the wall (Fig. \ref{fig:loop}c). Baffles break the spiraling
flow along the wall of the cylinder, which is a major feature of the
bifurcated state velocity field. The hysteretic cycle splits into two
classical first order cycles: the central symmetric state becomes stable
and can be obtained from any initial condition.

\paragraph*{Stability of the central branch (s).}

We focus now on the transition from symmetric state to bifurcated state
for curved blades without baffles. As mentioned before, the central
branch is very small and, for a given $(f, \; \theta)$, the
transition occurs after a certain time $t_{bif}$ which exhibits a complex statistics. 

\begin{figure}[!b] 
\begin{center}
\includegraphics[clip,width=6.cm]{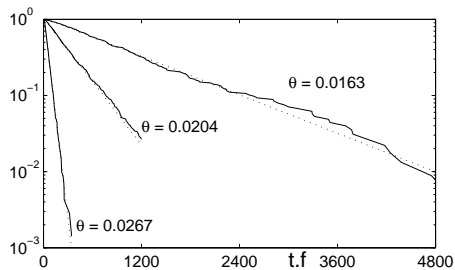}

\end{center} 

\caption{Cumulative density function of bifurcation times for three
different $\theta$ at $f=4.16\;$Hz. Dotted line: non-linear exponential fit. 
}

\label{fig:cdf} 
\end{figure}

\begin{figure}[!b] 
\begin{center}
\includegraphics[clip,width=6.5cm]{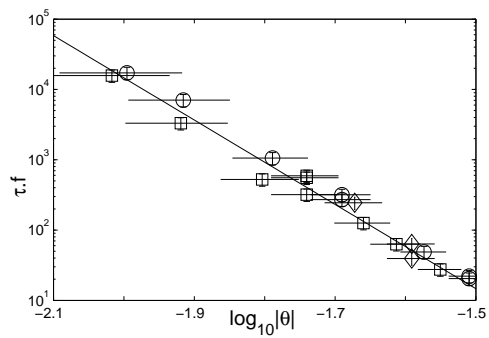}
\end{center} 
\caption{$\tau.f$ {\it vs.} $\theta$ for $f=4.16$ Hz/$Re= 3.3 \; 10^5$ ($\bigcirc$), $f=6$ Hz/$Re= 4.7 \; 10^5$ ($\Box$) and $f=10$ Hz/$Re= 7.9 \; 10^5$ ($\Diamond$), fitted by a $-6$ slope power law.}
\label{fig:taudetheta} 
\end{figure}

So we performed the following experiments: starting from rest, we
simultaneously start both motors to a given $(f, \; \theta)$ with a
short ramp (typically $1$ s) and record the torques. Few seconds after
the instant $t_{bif}$ when bifurcation occurs, we stop the motors, wait
a minute and run again. We perform typically $500$ runs to get the distribution of
bifurcation times. The cumulative distribution function (CDF) for $t_{bif}$
(Fig. \ref{fig:cdf}) shows exponential behavior for the probability of
staying in the symmetric state a time greater than $t$: $P(t_{bif}>t) =
A \; exp [-(t-t_{0})/\tau]$, $t_{0}$ is characteristic of the
transition duration ($t_0.f \sim 5$). Thus, we obtain a characteristic
bifurcation time $\tau (f,\theta)$ by non-linear fitting of the CDF.
We performed the experiment for three values of $f$. The results are shown in
Fig. \ref{fig:taudetheta} in log-log scale. There is no noticeable
dependence on $f$ and $\tau$ behaves as $\mid{\theta}\mid ^{-6}$.
So, as $\theta$ tends to zero, $\tau$ diverges very fast to infinity: the central
point is marginally stable. The physical phenomenon at the origin of such an exponent remains to be understood.

\paragraph*{Discussion.}

The experiment presented here opens mainly two problems: (i) the
existence and the nature of multiple regimes for this turbulent flow,
and (ii) the role of the noise or the fluctuations in some transitions
between these flow regimes, i.e., the stability problem for the
two-cell ($s$) branch.


We first try to explain the existence of multiple stable regimes by
hydrodynamical basic arguments. The von K{\'a}rm{\'a}n (VK) class of
Navier-Stokes solutions in semi-infinite space with one or two infinite
rotating disks for end-conditions has been extensively studied since
1921 \cite{karman21,batch51,stew53}. Experiments are necessarily limited
in diameter and do not strictly belong to the same class. However the
approximation is very commonly made at least for small $H/R$. In
practice, in our system, and in the spirit of Batchelor \cite{batch51}
and Stewartson \cite{stew53}, we construct finite-aspect-ratio solutions
of our experimental VK problem at high $Re$ (Fig.~\ref{fig:ldv1}) with
(i) any typical truncated Batchelor \cite{batch51} solution for $0 \leq
r \alt R/2$ together with (ii) some recirculation flow in rotation in
$R/2 \alt r \alt R$ and (iii) a thin boundary layer near the outer
cylinder which matches this rotation \cite{note:VKR}. The two-cell mean
flow ($s$) is simply described in the laboratory frame by two rotating
regions inertially driven by the blades and separated by a shear layer
near mid-height. Both disks centrifugally expel the fluid. Let's now
consider one-cell flows ($b_1$) and ($b_2$). Since one disk expels the
fluid and the other reinjects it to the center, these flows resemble the
corotating ($f_1.f_2<0$) regime solutions \cite{batch51,stew53}
characterized by: uniform rotation of the bulk; a boundary layer on each
disk; pumping from one disk to the other and recirculation at infinity.
This solution has no shear layer. Let's note that mean bulk rotation
---for $r \alt R/2$--- is close to zero (Fig.~\ref{fig:ldv1}, right). In
conclusion, we can make the assumption that flows ($b_1$) and ($b_2$)
are equivalent to corotating flows observed in two oppositely rotating
frames of frequencies $+f_r$ and $-f_r$ with $|f_r|> {\rm
max}(|f_1|,|f_2|)$. This is well consistent with the fact that the
single cell flows ($b_1$) and ($b_2$) exhibit global rotation in the
outer shell $R/2 \alt r \alt R$. The stability of such solution is
clearly enhanced by the concave curved blades that enforce rotation of
the fluid near the outer cylinder.


Let's now consider how these three solution branches exchange
their stability. First note that the bifurcation diagrams respect the
$\cal R_{\pi}$ symmetry: $\theta \rightarrow -\theta; \Delta K_p
\rightarrow -\Delta K_p$. The straight blades diagram
(Fig.~\ref{fig:loop}a) is continuous: from left to right two second
order transitions ($b_1$)$\leftrightarrow$($s$) and ($s$)$\leftrightarrow$($b_2$) 
are observed as in small $H/R$ systems \cite{dijhei83}. On the
contrary, the curved blades diagram (Fig.~\ref{fig:loop}b) is strongly
hysteretic. Addition of baffles (Fig.~\ref{fig:loop}c) allows to remove
a degeneracy: baffles drag disturb the outer cylinder boundary layer flow, thus
lowering the relative stability of one-cell flows with respect to the
two-cell flow. The large hysteresis cycle is split into two classical
first-order bifurcations. This singular cycle can thus be viewed as the
result of the collapse or collision of two first-order cycles. Similar
cycles are encountered in conical \cite{shthus96} and delta-wing flows \cite{gomzak85}.
The memory effect ---if the system is currently on ($s$), both
driving frequencies \emph{must} have been increased in parallel--- is thus essentially a
consequence of the cycle structure.


In order to test the effect of turbulence on the stability of the
observed flows, we lowered Reynolds number down to laminar using
water/glycerol mixtures. While $Re \alt 1000$, no multiplicity is
observed: the bifurcation diagram is similar to the straight-blade
diagram of Fig.~\ref{fig:loop}a. The cycle appears for $Re$ between 1000
and 3000. The study is in progress and will be reported elsewhere. The
high Reynolds behavior reported in this Letter is well established once
$Re \agt 5000$. Thus, multiplicity appears with turbulence and does not
with laminar ($Re \alt 110$) nor chaotic ($Re \alt 1000$) flows. A
possible explanation for the multiplicity could thus be the evolution of
the outer cylinder boundary layer with $Re$.


Besides, the statistical nature of the transitions themselves is
probably related to turbulent fluctuations. Let's first notice that the
bifurcation studied here corresponds to exchange of stability between
mean flows, these mean states being never realized at any given time. Is the bifurcation formalism exactly valid for our mean flows ? 
On some aspects, our system behaves as a low-dimension dynamical system, as
in the turbulent spiral transition observed in wide-aspect-ratio Taylor-Couette flow \cite{prigre02} or in the
noise-induced Hopf bifurcation for a Duffing oscillator with
multiplicative white noise \cite{malmar03}.
However, suppose a non-linear amplitude equation could correctly describe the
shape bifurcation diagram, it would probably not be able to catch the
statistics of the transition from the two-cell state ($s$) to a one-cell
state ($b_1$) or ($b_2$). This transition shows a very
peculiar statistics, with a very high critical exponent 6
(Fig.~\ref{fig:taudetheta}). It also strictly respects a
forbidden-transition rule: the horizontal axis of the bifurcation
diagram is never crossed, i.e., ($s$)$\rightarrow$($b_2$) [resp.
($s$)$\rightarrow$($b_1$)] is forbidden for $\theta<0$ [resp.
$\theta>0$]. This observational fact could by itself justify the
stability of the central point $\theta=0$, which has to respect both
rules. Furthermore, the non-crossing of the axis could be the signature
of multiplicative noise as suggested to account for small-scale
turbulence \cite{lavdub01}.


The global bifurcation reported in this Letter presents a very unusual
bifurcation diagram. Some features about the multistability have been
searched among the mechanics of high-Reynolds-number flows, while some
other simply involve the theory of non-linear bifurcations, possibly in
the presence of noise. Among the transitions, the two-cell $\rightarrow$
one-cell stability exchange plays a remarkable role, in presenting an
original statistics of transition and putting the flow definitively in a
state which breaks the $\cal R_{\pi}$ symmetry of the system and does not allow
the flow to restore statistically this symmetry when $Re \rightarrow
\infty$.


We thank V.~Padilla and C.~Gasquet for efficient assistance in building
and piloting the experiment, and B.~Dubrulle, O.~Dauchot and
N.~Leprovost for fruitful discussions.

\end{document}